# Effectiveness of Common Fabrics to Block Aqueous Aerosols of Virus-like Nanoparticles

AUTHOR NAMES: *Steven R. Lustig\*, John J.S. Biswakarma, Devyesh Rana, Susan H. Tilford, Weike Hu, Ming Su, Michael S. Rosenblatt*

AUTHOR ADDRESS: Steven R. Lustig – Department of Chemical Engineering, Northeastern University, Boston, Massachusetts 02703, United States; Email: s.lustig@northeastern.edu

ABSTRACT: Layered systems of commonly available fabric materials can be used by the public and healthcare providers in face masks to reduce the risk of inhaling viruses with protection about equivalent or better than the filtration and adsorption offered by 5-layer N95 respirators. Over 70 different common fabric combinations and masks were evaluated under steady state, forced convection air flux with pulsed aerosols that simulate forceful respiration. The aerosols contain fluorescent virus-like nanoparticles to track transmission through materials that greatly assist the accuracy of detection, thus avoiding artifacts including pore flooding and the loss of aerosol due to evaporation and droplet break-up. Effective materials comprise both absorbent, hydrophilic layers and barrier, hydrophobic layers. Although the hydrophobic layers can adhere virus-like nanoparticles, they may also repel droplets from adjacent absorbent layers and prevent wicking transport across the fabric system. Effective designs are noted with absorbent layers comprising terry cloth towel, quilting cotton and flannel. Effective designs are noted with barrier layers comprising non-woven polypropylene, polyester and polyaramid.

TABLE OF CONTENTS GRAPHIC:

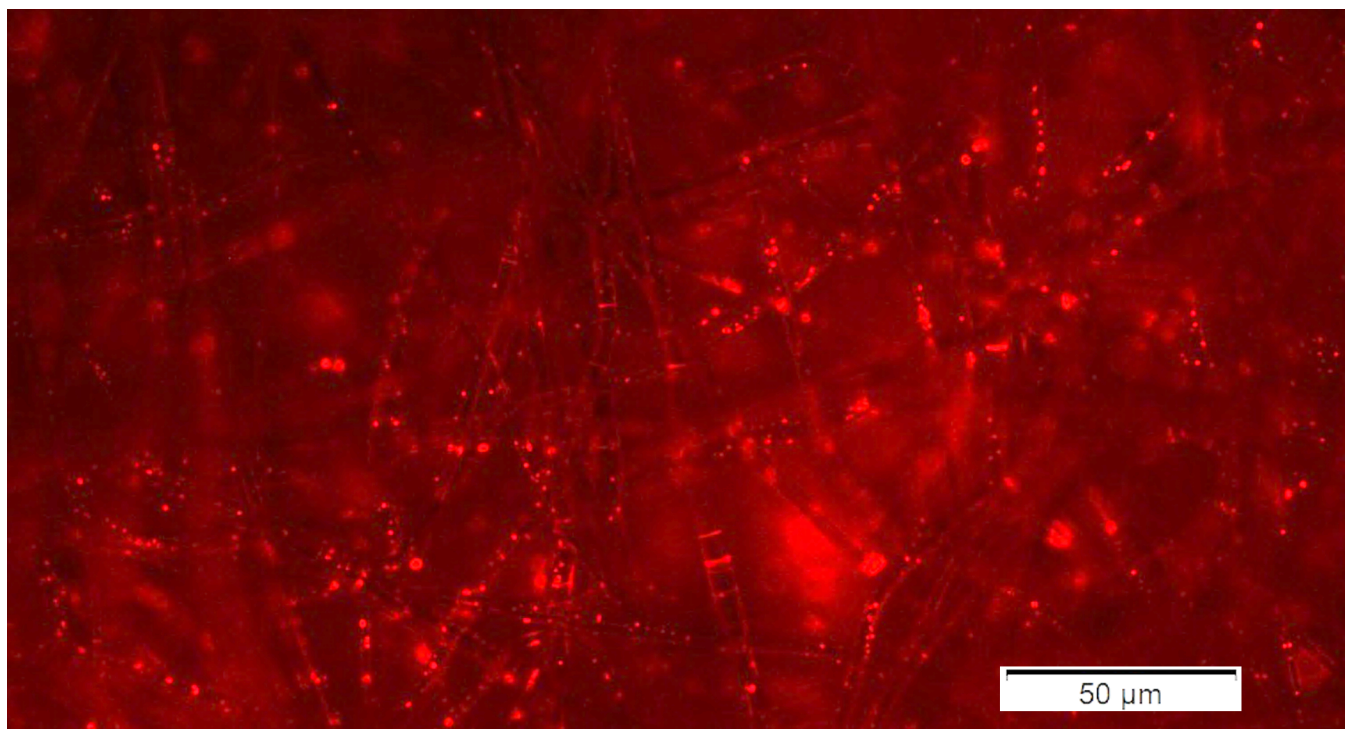

Fluorescent virus-like nanoparticles caught from pulsed aerosols onto the surface of a nonwoven polyester fiber.

The personal protective equipment (PPE) shortage in the United States during the SARS-CoV-2 pandemic has put the issue of its availability directly into the public domain. Healthcare PPE shortages in the U.S. were triggered by massive shipments to China in early 2020[1,2] and subsequent production and transportation stoppages from outbreaks in China and Southeast Asia, source of 80-90% of the U.S. PPE supply.[3] Masks reduce inhalation of aqueous viral aerosols emitted from infected individuals when talking, coughing or sneezing.[4-7] Masks may also be beneficial by serving as a reminder for wearers to avoid touching their face and thus prevent transmission from the hands to the user's nose, mouth, and eyes. Moreover, the CDC and state governments now either require or recommend the public wear face masks in public.[8] It is increasingly more urgent to identify effective fabrics and mask designs to the public so there is no competition for healthcare provider PPE.[9] Understanding effective mask construction may enable safe homemade masks and reduce the PPE supply issues during the pandemic. Moreover, the critical shortage of certified respirators and masks faced by Massachusetts hospitals forced hospital personnel to consider time-sensitive solutions for alternative PPE. Ideally, alternative PPE would be facile to assemble from largely available fabric stocks of local vendors and provide approximately equivalent or superior virus particle filtration as the certified PPE.

Commercially manufactured, certified respirators and surgical masks are generally considered more effective than homemade masks. N95 respirators that tightly seal around the mouth and nose are typically worn by healthcare providers caring for patients with infectious conditions that transmit via aerosolized pathogens. Surgical masks are designed to block direct fluid entry into the wearer's nose and mouth from a splash, cough, or sneeze; and are not designed to block aerosolized pathogens. Materials used for health care service face masks are subject to extensive performance criteria,[10] including bacterial filtration efficiency, particle filtration efficiency, fluid flow resistance, air flow resistance, flame propagation rate, and skin reactivity as mandated by the National Institute for Occupational Safety and Health (NIOSH).[11] The first two are directly related to the effectiveness of the material to serve as a barrier to aqueous viral aerosols. ASTM F2100-19E1 specifies assessing filtration using 100 nm sized particles of salt aerosol.[12] The N95 certification indicates that 95% of the total particles in a salt aerosol with an average particle size of 300nm are blocked by a material under standard conditions. N95 respirators are typically used by physicians and surgeons. ASTM F2299/F2299M-03(17) uses light scattering particle counting of latex spheres between 100 nm and 5 $\mu$m in diameter.[13] Bacterial efficiency standards are described in ASTM F2101 which requires aqueous bacterial aerosols having 3 $\mu$m diameter droplets.[14] Meanwhile, aqueous aerosols from speech, sneezing and coughing have size distributions spanning several orders of magnitude[7,15,16] up to thousands of microns. The SARS-CoV-2 virus itself is found in various shapes with polydisperse diameters ranging between 60 and 140 nm.[17] Salt particulates, latex spheres, bacteria and viruses are widely diverse in size, shape, surface chemistry and interfacial properties. These properties can affect the transport and adhesion within the complex surfaces of materials used in PPE face masks.

Comparative studies of common, household fabrics generally indicate these materials are more permeable than medical grade PPE and widely variable in their filtration efficiencies. For example, Rengasamy[18] provides caution that fabrics can exhibit a range of filtration efficiencies. Examples of sweatshirts, t-shirts, towels, and scarfs from different manufacturers were tested with polydisperse (75 ± 20 nm) salt aerosols and 13 sizes of monodisperse salt aerosols (20 nm-1 $\mu$m) at face velocities of 5.5 and 16 cm/s. Particulate transmission through the materials was determined by measuring the particle count upstream and downstream of the filter media using a scanning mobility particle sizer. Fractional transmissions ranged from 40-90% for the polydisperse aerosol and 40-97% for the monodisperse aerosols at 5.5 cm/s. Davies[19] investigated the filtration performance of common household fabrics to remove airborne viruses and bacteria. Fabrics were exposed to aerosols containing either *Bacillus atrophaeus* (0.95-1.25 $\mu$m) or Bacteriophage MS2 (23 nm). Aerosols were delivered in a closed chamber at 30 L/min and particle counts were measured upstream and downstream of the filter media. Based on a combination of filtration efficiency and pressure drop, the highest performing fabrics were 100% cotton t-shirt and pillowcase. The surgical mask had a 96% mean filtration efficiency for the 1 $\mu$m particles and 90% for the 23 nm particles. In comparison, the 100% cotton t-shirt had a 69% mean filtration efficiency for the 1 $\mu$m particles and 51% for the 23 nm particles. The pillowcase had a 61% mean filtration efficiency for the 1 $\mu$m particles and 57% for the 23 nm particles. Hence these fabrics are far more permeable than N95 respirators. These investigations did not attempt to combine multiple types of fabric layers to achieve comparable performance as the NIOSH certified medical respirators and masks, such as the N95 respirators. Recently, Konda[20] studied several common fabrics such as cotton, silk, chiffon, flannel, polyester blends with up to two layers. Cotton quilt and cotton/chiffon performed about as well as an N95 respirator at filtering saline aerosols. Although the methodology is in compliance with NIOSH 42 CFR Part 84 test protocol, the instruments are noted to have poorer counting efficiencies for particles smaller than about 300nm. Furthermore, unknown fractions of the aqueous aerosol particles are lost by evaporation as well as break-up into undetectable, smaller droplets. These aerosols did not contain virus nanoparticles that could be independently identified when transported through the materials. Blocking the transport of virus particles is a prime function of mask fabric.

In response to the time-sensitive need for alternative PPE, we identify commonly available fabric materials that the public and healthcare providers can use in face masks to reduce the risk of viral aerosol inhalation. Over 70 different common fabric multi-layer designs are compared to NIOSH certified medical respirators and ASTM certified masks for filtration efficiency using protocol conditions similar to those of ASTM standards. A common design theme emerges for many layered fabric designs that may reduce the risk of viral inhalation from aerosolized contamination directly striking the mask in both healthcare-patient interactions as well as public interactions with limited physical distancing.

RESULTS AND DISCUSSION

Fluorescent, virus-like nanoparticles emulate the size and surface character of SARS-CoV-2 virus particles and are readily detected and counted. Rhodamine 6G is incorporated into nanoparticles as it is highly photostable and fluoresces with high quantum yield efficiency. It remains well partitioned within the nanoparticle matrix of poly(lactic-co-glycolic acid). Figure 2 is a scanning electron microscope image of a small cluster of primary nanoparticles. Most of the encapsulated nanoparticles have spheroidal shape with some shallow wrinkles. Wrinkles may be due to the sheer stress present during the formation of the core-shell structure. The measured primary particle sizes of the nanoparticles range between 10 and 200 nm, which is the same range as SARS-CoV-2 virus particles[17], see Figure S1. Zeta potential measurements indicate neutral surface charge over six decades of concentration, see Figure S2. Detailed synthesis methodology and characterization results are provided in the Methods Section.

Transport of nanoparticles in aqueous aerosol is predicated on forced convection air flux. For example, placing the aerosolizer jet in direct contact with the surface of an N95 respirator at 20 kPa gauge pressure results in instantaneous surface accumulation of water. No nanoparticles were detected on the opposite side of the respirator. This is condition occurred for all materials and masks, except for the most open, highly porous weaves. Direct aerosol jetting onto the densely woven fabrics exhibits the same surface flooding result. Pore flooding traps nanoparticles, preventing transmission through the material. This result is independent of the aerosol pressure that could be applied. Similar pore flooding can occur in salt solution aerosol testing. Nanoparticle transmission through porous materials begins to occur without pore flooding as the steady-state volumetric flow rate of air exceeds the incident volumetric flow rate of aqueous aerosol. Partial flooding decreases the effective material porosity and leads to exaggerated filtration efficiency. In practical terms dense fabric masks do not transmit nanoparticles into a mask, such as virus particles, without active respiration or permeating air convection.

The transmission measurement of nanoparticles through mask materials is based on test conditions that emulate ASTM methods, enable high precision and repeatability, and reproduce sensibly physiological conditions. The rate of human ventilation at rest is nominally 6 L/min[21] and can increase several-fold upon active exertion. Our testing establishes a baseline steady-state air flow of 14 L/min through each test material. Each test is subjected to a total threat of 2 mL aqueous solution containing the fluorescent virus-like nanoparticles at 0.5 mg/mL. This total threat volume is delivered by 26 pulses of aerosol, each lasting one second. The duration and overpressure of the pulses emulates forceful expiration, i.e. a spray resulting from a sneeze, cough or speech from an infected individual. The steady-state air flow being in excess of restful ventilation replicates a slightly elevated ventilation rate as a safety margin, prevents pore flooding, and enables improved statistical repeatability in the nanoparticle count measurements. The pulsed aerosol droplets are polydisperse in size and closely match the size range from forceful expiration. Nanoparticles transmitted through the test material are collected at a distance of 1 mm on a glass slide. The gas flow and slide placement configure the system to be well within the estimated collection regime, i.e. particle capture limit.[22] After a nanoparticle collides with the glass, the rebounded kinetic energy is insufficient to escape the attractive potential energy. Specific details about the aerosol transmission testing are described in the Methods Section, also see Figure 1.

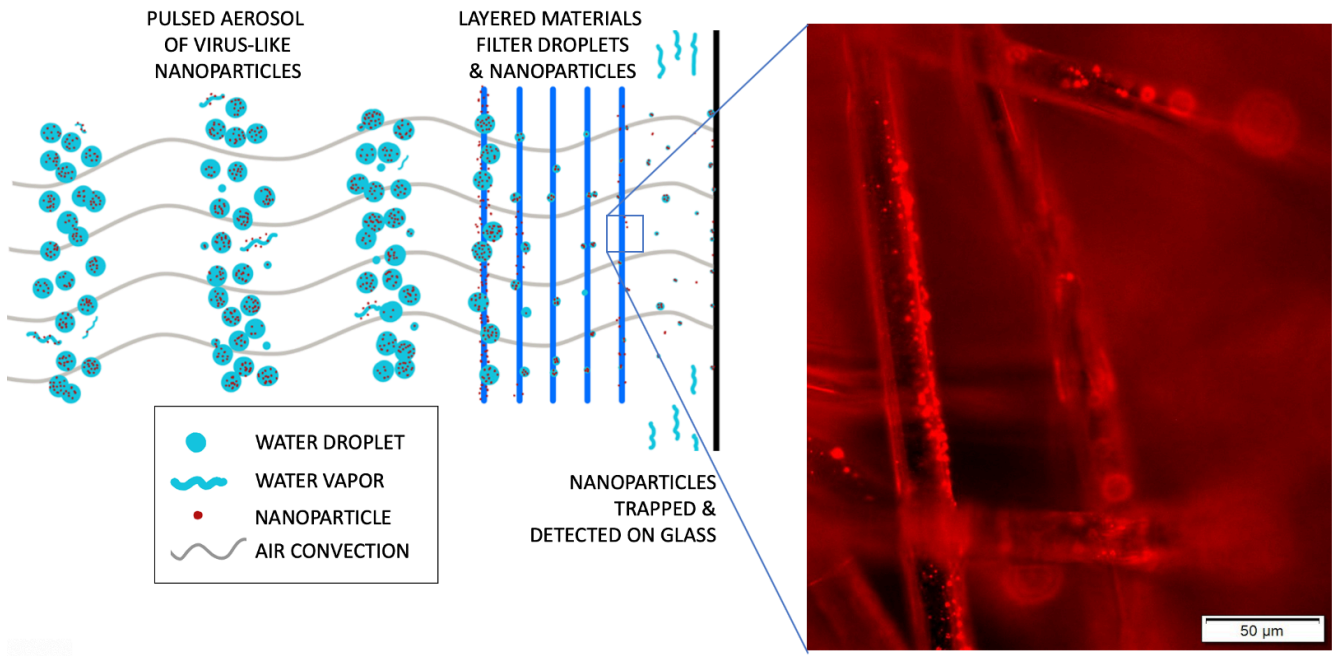

**Figure 1.** Schematic drawing of pulsed aqueous aerosol containing fluorescent, virus-like nanoparticles being drawn through layered materials by steady-state, forced convection air flux until transmitted nanoparticles are collected on glass slide (Left). Representative fluorescent micrograph of fluorescent, virus-like nanoparticles trapped on non-woven polypropylene material (Right). This illustration was created by Shoshanna Lustig for this article.

Over 70 different common material arrays were evaluated under steady-state air permeation against pulsed aerosols that simulate forceful expiration. A list of materials is provided in the Supporting Information, see Table S1. Table 1 summarizes our most notable transmission results and comparative statistics. Data for the 5-layer N95 respirator by 3M™ are provided in the first row. This is the standard PPE recommended by the Centers for Disease Control and Prevention (CDC) when caring for SARS-CoV-2 patients undergoing an aerosolizing procedure. Several 30 mm diameter samples were cut from around the respirator. The fractional transmission is the nanoparticle count transmitted through the material, normalized by the incident nanoparticle count. The fractional transmission standard deviation across all sampled locations exceeds the typical standard deviation for the nanoparticle counting measurement. This suggests that the filtration efficiency is dependent on the location of the mask. This is reasonable for a stack of non-woven layers that are pressed heterogeneously into a shape comprising highly varying curvature and thickness. Nonetheless the overall average and standard deviation of nanoparticle counts over 69 independent measurements are provided across the entire respirator. Specifically, the 5-layer N95 respirator by 3M™ displayed a fractional transmission of 0.56 ± 0.30 ppt. Remaining materials shown in Table 1 exhibited uniform fractional transmission among multiple replicate samples. Standard surgical masks, also evaluated in this study, are currently recommended by the CDC while caring for SARS-CoV-2 patients not undergoing an aerosolizing procedure. There are three performance classifications for the remaining materials based on the normalized permeability index, i.e. the fractional transmission of the material divided by the fractional transmission of the 5-layer N95 respirator. Thus, the permeability index for the 5-layer N95 respirator is unity. This index is included with the *p*-value indicating that the fractional transmission of the material is indistinguishable from the fractional transmission of the 5-layer N95 respirator. Here $p < 0.05$ represents 95% confidence that the two materials are distinguishable. This is a double tailed test because materials may be distinguishable by having significantly higher fractional transmission or lower fractional transmission than the 5-layer N95 respirator. The *p*-value is computed using Welch's t-test as the variances of the material and 5-layer N95 respirator are unequal and must be estimated separately.

Several layered systems exhibit fractional transmission statistically lower than or equivalent to the 5-layer N95 respirator. Specifically, Sheldon G mask with cellulose filter and combination masks, combining two outer layers of white denim with two inner layers of OLY-FUN nonwoven polypropylene and two-layer of Kona quilting cotton with four layers of OLY-FUN exhibit fractional transmissions of

0.16 ± 0.06 ppt, 0.31 ± 0.07 ppt, and 0.40 ± 0.18 ppt, respectively. These mask designs achieve 72%, 55%, and 28% lower fractional transmission than the 5-layer N95 respirator, respectively. Effective materials comprise both absorbent, hydrophilic layers and barrier, hydrophobic layers. Although the hydrophobic layers can adhere virus-like nanoparticles, they may also repel droplets from adjacent absorbent layers and prevent wicking transport. High fiber density and tortuosity increases the probability of collision with aerosol droplets. Effective designs are noted with absorbent layers comprising terry cloth towel, quilting cotton and flannel. For example, two layers of terry cloth, two layers of white flannel, and four layers of Kona quilting cotton exhibit fractional transmissions of 0.50 ± 0.12, 0.51 ± 0.24, and 0.62 ± 0.17, respectively. These commonly available mask materials exhibit fractional transmissions within 10% of the five-layer N95 respirator. Effective designs are noted with barrier layers comprising OLY-FUN (non-woven polypropylene), lab coat (polyester/polyaramid), cotton coated with spray-on fabric protector as well as traditional synthetic aliphatic and aromatic polymer fibers. Although some terry cloth and cotton multi-layers are effective alone, inclusion of an additional hydrophobic repelling layer is recommended to prevent wicking transport for higher volume threats. Sole use of denim is not effective- in general the yarn bundles are very dense but spaced with wide interweave gaps to promote breathability in jeans. This is demonstrated by high fractional transmission by two layers each of 4oz light weight blue denim, 7oz midweight blue denim, and 11oz heavy weight stretch black denim of 3.91 ± 1.82 ppt, 7.61 ± 0.63 ppt, and 9.43 ± 0.99 ppt, respectively. The two-layer denims exhibit 698%, 1359%, and 1684% higher fractional transmission than the 5-layer N95 respirator, respectively. The fusible polyesters considered are also highly porous. Several additional layered systems exhibit fractional transmission statistically equivalent to the duckbill surgical mask. These may be effective in conjunction with additional safeguards, such as social distancing, and smaller threat volumes.

**Table 1.** Permeability of barriers tested featuring commonly available fabrics and materials. Fractional transmission is the nanoparticle count transmitted normalized by the incident nanoparticle count, reported with number of independent particle count measurements, N. N95 Normalized Permeability Index is the fractional transmission of the material divided by the fractional transmission of the N95 Mask (first table entry), reported with the unequal variances t-test probability that the transmission is no different from the N95 Mask. Notes: (a)Average of data collected from multiple positions around the mask. The data show indications that the transmission is dependent on location on the pressed mask. (b) Design of Sheldon Gentling: outer most layer comprises: ProCool Stretch-FIT Dri-QWick Sports Jersey Fabric by AKAS Textiles & Laminations, inner most layer comprises Zorb 3D Stay Dry Dimple Heavy Duty Fabric by AKAS Textiles & Laminations. All materials supplied by Wazoodle Fabrics. (c) Kona® quilting cotton fabric, supplied by Joann Fabrics, Hudson, Ohio (d) 65 GSM (Gram per square meter) polypropylene nonwoven fabric. (e) Lab coat is a blend of polyester and polyaramid. (f) Pellon® Midweight #931TD fusible polyester. (g) Pellon® #SF101 fusible polyester. (h) Kona® quilting cotton fabric treated with 2.2wt% Scotchgard™. (i) Pellon® #P44F fusible polyester. (j) https://www.joann.com/how-to-make-a-denim-face-mask/042188731P326.html (accessed Apr 21, 2020). (k) HTC (High Thread Count), 525 horizontal and vertical thread counts/inch.

| Material | Fractional Transmission, parts per 1000 (N) | N95 Normalized Permeability Index (*p*-value) |
|---|---|---|
| N95 Mask (3M™: #1860S Lot #15886, 5 layer) | 0.56 ± 0.30 (69)[a] | 1.0 (---) |
| **Transmission Statistically Lower Than N95 Mask ($p < 0.05$)** | | |
| Sheldon G mask with cellulose filter[b] | 0.16 ± 0.06 (27) | 0.3 (0.001) |
| White Denim/ OLY-FUN (x2)/ White Denim | 0.31 ± 0.07 (9) | 0.5 (0.001) |
| Kona cotton[c]/ OLY-FUN[d] (x4)/ Kona cotton | 0.40 ± 0.18 (18) | 0.7 (0.004) |

| **Transmission Equivalent to N95 Mask** ($p > 0.05$) | | |
|---|---|---|
| N95 Mask (3M™: #8200 Lot #B18198, 3 layer) | 0.47 ± 0.11 (36) | 0.8 (0.148) |
| Kona cotton (x2)/ Terry Cloth (x2) | 0.50 ± 0.18 (18) | 0.9 (0.232) |
| Terry cloth towel (x2) | 0.50 ± 0.12 (18) | 0.9 (0.145) |
| Kona cotton (x4) | 0.51 ± 0.24 (9) | 0.9 (0.514) |
| Lab Coat[e]/ Flannel/ OLY-FUN (x2)/ Kona cotton | 0.57 ± 0.26 (9) | 1.0 (0.942) |
| Kona cotton/Flannel /OLY-FUN (x2)/ Kona cotton | 0.62 ± 0.06 (18) | 1.1 (0.116) |
| White flannel (x2) | 0.62 ± 0.17 (18) | 1.1 (0.318) |
| Heavy Tee Shirt 100% cotton (x2) | 0.64 ± 0.06 (18) | 1.1 (0.060) |
| Lab Coat (x2)/ Flannel (x2) | 0.69 ± 0.20 (9) | 1.2 (0.093) |
| White 12oz denim/ Kona cotton (x2)/White 12oz denim | 0.70 ± 0.23 (9) | 1.2 (0.122) |
| Kona cotton/ White 12oz denim (x2)/ Kona cotton | 0.79 ± 0.62 (9) | 1.4 (0.293) |
| Kona cotton/ OLY-FUN (x2)/ Kona cotton | 1.10 ± 0.89 (9) | 2.0 (0.072) |
| **Transmission Statistically Higher than N95 Mask** ($p < 0.05$) | | |
| Procedure cone mask (Cardinal Health™, #AT7509) | 0.68 ± 0.08 (18) | 1.2 (0.003) |
| Terry cloth towel (x1) | 0.73 ± 0.14 (9) | 1.3 (0.005) |
| Kona cotton/ White flannel/ Kona cotton | 0.73 ± 0.05 (18) | 1.3 (0.001) |
| Kona cotton (x3) | 0.85 ± 0.15 (9) | 1.5 (0.001) |
| Kona cotton/ Pellon Midweight[f] | 0.86 ± 0.23 (72) | 1.5 (0.001) |
| KN95 mask (GB2626-2006KN95) | 0.91 ± 0.24 (18) | 1.6 (0.001) |
| Kona cotton (x2) | 0.92 ± 0.05 (18) | 1.6 (0.001) |
| Kona cotton/ Pellon[g]/ Kona cotton | 0.95 ± 0.33 (45) | 1.7 (0.001) |
| Duck bill surgical mask (Halyard™ #37525) | 0.98 ± 0.37 (18) | 1.7 (0.001) |
| Kona cotton/ Kona 2.2wt% Scotchgard[h]/ Kona cotton | 1.01 ± 0.20 (18) | 1.8 (0.001) |
| Kona cotton/ Polartec®/ Kona cotton | 1.04 ± 0.38 (18) | 1.8 (0.001) |
| White Flannel (x1) | 1.04 ± 0.08 (18) | 1.8 (0.001) |
| Heavy Tee Shirt 100% cotton (x1) | 1.07 ± 0.10 (18) | 1.9 (0.001) |
| Kona cotton/ Pellon[i]/ Kona cotton | 1.14 ± 0.60 (9) | 2.0 (0.004) |
| White 12oz denim/ Pelon[f]/ White 12oz denim[j] | 1.22 ± 0.77 (27) | 2.2 (0.001) |
| Kona cotton/ White 12oz denim/ Kona cotton | 1.42 ± 0.51 (9) | 2.5 (0.001) |
| HTC[k] pillowcase/ Flannel/ OLY-FUN (x2)/HTC pillowcase | 1.47 ± 0.66 (9) | 2.6 (0.001) |
| OLY-FUN polypropylene nonwoven 65GSM (x2) | 2.56 ± 0.74 (9) | 4.5 (0.001) |
| 4oz Light weight blue denim (x2) | 3.91 ± 1.82 (9) | 6.9 (0.001) |
| 7oz Midweight blue denim (x2) | 7.61 ± 0.63 (5) | 13.5 (0.001) |
| 11oz Heavy weight stretch black denim (x2) | 9.43 ± 0.99 (18) | 16.7 (0.001) |

CONCLUSIONS

Commonly available fabric materials can be used by the public and healthcare providers in face masks to reduce the risk of inhaling viruses from aerosols generated by coughs, sneezes and speech from infected individuals. The protection by some layered designs offer protection about equivalent or better than the filtration and adsorption offered by 5-layer N95 masks. Effective materials comprise both absorbent, hydrophilic layers and barrier, hydrophobic layers. Although the hydrophobic layers can adhere virus-like nanoparticles, they may also repel droplets from adjacent absorbent layers and prevent wicking transport. Effective designs are noted with absorbent layers comprising terry cloth towel, quilting cotton and flannel. Effective designs are noted with barrier layers comprising non-woven polypropylene, polyester, polyaramid.

This work responds to the time-sensitive need for alternative personal protective equipment for healthcare workers as well as face masks for the public. Considering the results of this work and prior work, recommended mask designs include those multi-layered combinations in Table 1 that exhibit transmission either equivalent or lower than the transmission offered by 5-layer N95 masks. It is critical that the materials edges conform snugly to the face to prevent aerosol from entering gaps between the face and mask. The mask must not enable viral imbibition by the lips, tongue and saliva. Ideally the mask does not contact the lips, or there is at least one hydrophobic layer fabric in contact with the face, so aerosol trapped from the exterior does not wick through the mask and become transported by the mouth. Since aerosol transport through a mask is predicated on forced convection air flux, it is recommended that individuals wearing masks reduce inhalation intensity when placed in contact of an unsafe aerosol.

## METHODS

### Virus-Simulant Nanoparticles

Materials:

Ethyl acetate, poly(lactic-co-glycolic acid) (PLGA), eicosane, rhodamine 6G, and polyvinyl alcohol (PVA) were purchased from Sigma Aldrich (Billerica, MA, USA) and were used as is without any further processing or purification.

PLGA Nanoparticle Preparation:

Nanoparticles (NP) were prepared by mixing 100 mg PLGA pellets with 1 mL ethyl acetate, 20 µg rhodamine 6G, and 12 mg eicosane. The resulting mixture was vortexed for 5-10 minutes until homogenized. 2 mL of 5 wt% PVA was added and sonicated for 2 minutes using an ice water bath to prevent evaporation of ethyl acetate. This solution was mixed to 50 mL of 3 wt% PVA solution immediately after sonication and stirred at 800 RPM for 2 hours until the ethyl acetate evaporated. The resulting solution was split into two centrifuge tubes and centrifuged at 6000 RPM for 5 minutes followed by the removal of the supernatant. The remaining precipitate was diluted with deionized water and vortexed for another 5 minutes. The centrifugation and rinse were repeated 3 times. The final precipitate was diluted with 30 mL water to obtain a final experimental concentration of ca. 7 mg/mL. A small aliquot of dispersion was weighed both wet and dry to determine accurately the actual NP concentration. This stock solution was further diluted to 0.5 mg/mL for experimentation. This concentration was chosen after a series of experiments to determine optimal NP concentration such that NP do not aggregate, did not clog the fabrics and did not clog the aerosol generator.

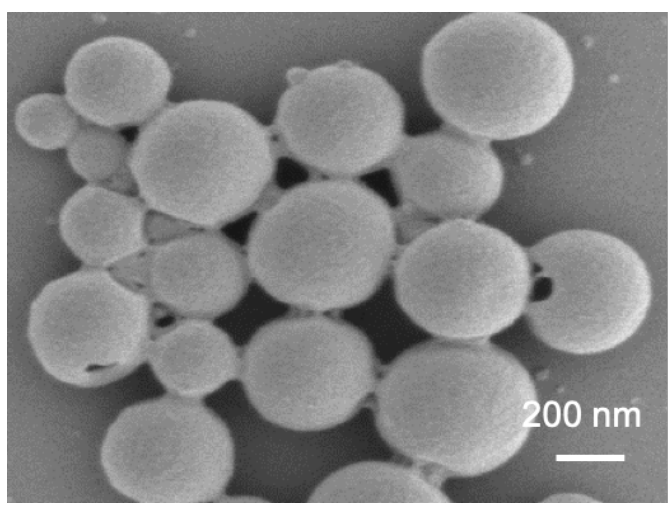

**Figure 2.** Scanning electron micrograph (SEM, Ultra 55, 10 keV) of a small cluster of primary nanoparticles. The core-shell structure is not thermally stable under the exposure to high energy density, such as a focused electron beam in higher magnification and the nanoparticle will partially melt to present irregular shape.

PLGA NP Size Distribution and Zeta Potential Tests:

NP size distribution, shown in Figure S1, and zeta potential tests, shown in Figure S2, were conducted using a Malvern Zetasizer Nano ZS90 and the accompanying Malvern Zetasizer v7.12 software. Polystyrol/Polystyrene (D-51588) cuvettes from Sarstedt were used for sample loading and measurements. The stock solution concentration of NP of 7.0 mg/mL (or 1X) was serially diluted to

achieve 10X, 50X, 100X, 200X, 400X, 800X, 1600X, 3200X, 6400X, 12800X, 25600X, and 51200X dilution factors. Exactly 1 mL of the diluted solutions were loaded into a cuvette and placed within the Zetasizer instrument. For each dilution, 3 samples were prepared, and 3 measurements were taken per sample (n = 9) using a 173° backscatter measurement angle. The Zetasizer was configured for size measurements using PLGA@eicosane with a refractive index of 1.570 and absorption value of 0.001 with a dispersant of water at 25°C. For NP size measurements, no other settings were required whereas for zeta potential measurements, a Smoluchowski model is applied with an $F(\kappa a) = 1.50$, where $\kappa$ is the Debye length and $a$ is the radius of the particle. The NP size distribution and zeta potential were then plotted using Graphpad Prism v8.0.0.

Aerosol Transmission Testing

Test apparatus:

A test apparatus was designed to analyze the degree of transmission of aerosols through various materials. Figure 3 illustrates a schematic of the test apparatus and identifies the components. A labeled photograph shows the actual components in Figure S3. Design parameters for this system were informed by ASTM procedures that involve testing the performance of surgical masks in filtering aerosols.[23-25] The Master Airbrush Pro Gravity Feed Airbrushing System ECO KIT-17 is used to generate an aerosol containing the fluorescent nanoparticle solution as shown in Figure S3. A Master air compressor TC-20 pressurizes the solution to 20 kPa. The pressurized solution is emitted from the Master airbrush G22 as an aerosol due to shearing interactions at the airbrush tip with an opening diameter of 345 $\mu$m. For each trial, 2 mL nanoparticle solution is emitted from the airbrush in bursts with a duration of one second every five seconds until the airbrush fluid tank is depleted. The aerosol is released into a 1L vacuum filter reservoir sealed over a glass bottle during a steady state 14 L/min volumetric flow of air set using Sho-Rate rotameter #012. The vacuum filter is sealed so that the volumetric flow rate is approximately uniform within the test apparatus, and it is controlled so that the contained fluids exhibit laminar flow (Re = 1900 < 2000). The velocity of the aerosol at the nozzle facing the material samples estimated to be 297 cm/s. For each material a 30 mm diameter sample is cut and held tightly with an O-ring over a nozzle with an inner diameter of 10 mm. The material samples are held taut, and all samples consisting of layered materials are necessarily held without spacing between adjacent layers. As shown in Figure S3, a 0.5" x 0.5" glass slide is positioned 1 mm from the material sample to collect aerosol and droplets that are transmitted. A circle drawn on the opposite face of the glass slide indicates the position of the slide that aligns with the center of the material sample, and the aerosol that accumulates on the side facing the sample is analyzed using fluorescence microscopy.

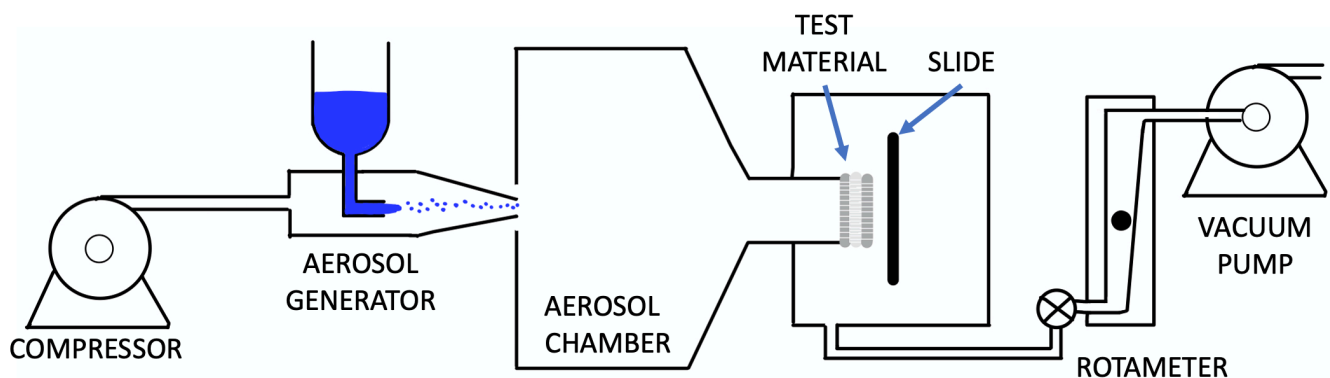

**Figure 3.** Schematic drawing of test apparatus. An air brush comprises the compressor and aerosol generator in which virus-like nanoparticles are dispersed in solution (blue) and gravity fed into the forced convection air flux that is mediated by a trigger (not shown) to create pulsed aerosol sprays. Aerosol is immediately sprayed into a 1L chamber leading to a nozzle capped by the test material (gray layers). A glass slide (thick black) captures nanoparticles transmitted from the right edge of the test material, while air flow proceeds through a needle valve, rotameter and steady vacuum pump.

Aerosol Droplet Size Distribution:

Droplet size distribution was determined by using the spray apparatus and spraying directly onto a 0.5" x 0.5" glass slide. The spray collected from one aerosol burst was then evaluated under a Keyence VHX-970F Optical Microscope from Keyence Corporation (Itasca, IL, USA). Images were captured at 20X magnification for large droplets and aerosols and 100X magnification for all droplets to understand the full droplet size distributions. A total of 64 images were taken. The raw images were further processed using ImageJ[26] to subtract the background with a 50 pixel rolling ball radius and a dark background. A scale of 26 pixels was identified as the equivalent of 10 μm. The images were also cropped from the bottom by 50 pixels to remove the magnification and scale bar texts to remove any erroneous particles being counted due to the text. The image was then converted to an 8-bit image format to which a minimum and maximum contrast threshold was set to 0 and 225, respectively. This resulted in black (droplets) and white (background) images. These black and white images were then counted and measured using the counting function of ImageJ, within the Analyze feature, using an ellipse outline method. An example of the subsequent image alterations is located in Figure S4. Figure S5, plots the counted ellipses and measured diameters in a total distribution of droplets by size and frequency. The aforementioned procedure was automated by creating a custom Plugin using ImageJ's batch scripting language, to remove human bias during image analysis and to speed up analysis. The veracity of the script was confirmed by manual analysis of each step, per the image output examples. The size distribution and frequency were then plotted using Graphpad Prism v8.0.0.

Nanoparticle Distribution After Transmission Through Fabrics:

Nanoparticle distribution was measured by placing 1 cm x 1 cm glass slides onto the glass holder within the test apparatus and sprayed with fluorescent rhodamine tagged PLGA NP. The NP containing glass slides were then observed under a fluorescent microscope using an Olympus BX43 Fluorescent

Microscope, containing an Olympus U-TV1XC center and Olympus XM10 camera. An X-CITE 120LED Boost laser controller from Excelitas Technology was used for a fluorescent laser source run at 45% power for fluorophore excitation. At least 9 images were taken at 20X optical zoom per fabric to determine particle concentration per area and multiple experiments were conducted per fabric using the accompanying Olympus cellSense Standard 1.16 software. A constant gain and exposure were chosen of 18 dB and 1.109 s, respectively, and a fixed scale contrast was applied between 0 and 5000. Individual images were post-processed in ImageJ, similar to the Droplet Size Distribution protocol. The raw images were further processed using ImageJ to subtract the background with a 500 pixel rolling ball radius and a dark background. A scale of 160 pixels was identified as the equivalent of 50 $\mu$m. The images were cropped from the bottom by 50 pixels to remove the magnification and scale bar texts to remove any erroneous particles being counted due to the text. The image was then converted to an 8-bit image format to which a minimum and maximum contrast threshold was set to 15 and 250, respectively. This resulted in black (droplets) and white (background) images. These black and white images were then counted and measured via the ImageJ counting feature, within the Analyze feature, using an ellipse outline method. An example of the subsequent image alterations is located in Figure S4. The ellipses are counted, and diameter measured to obtain the total distribution of droplets by size and frequency. The aforementioned procedure was automated by creating a custom Plugin using ImageJ batch scripting language, to remove human bias during image analysis and to exponentially speed up analysis. The veracity of the script was confirmed by manual analysis of each step. The nanoparticle count and size distribution are included in Table 1.

AUTHOR INFORMATION

Corresponding Author

Steven R. Lustig – Department of Chemical Engineering, Northeastern University, Boston, Massachusetts 02703, United States; Email: s.lustig@northeastern.edu

Authors

John J.S. Biswakarma – Department of Chemical Engineering, Northeastern University, Boston, Massachusetts 02703, United States

Devyesh Rana – Department of Chemical Engineering, Northeastern University, Boston, Massachusetts 02703, United States

Susan H. Tilford – Department of Chemical Engineering, Northeastern University, Boston, Massachusetts 02703, United States

Weike Hu – Department of Chemical Engineering, Northeastern University, Boston, Massachusetts 02703, United States

Ming Su – Department of Chemical Engineering, Northeastern University, Boston, Massachusetts 02703, United States

Michael S. Rosenblatt – Department of General Surgery, Lahey Hospital and Medical Center, 41 Mall Road, Burlington, Massachusetts 01805, United States

SUPPORTING INFORMATION AVAILABLE

Supporting information accompanying this article includes a description and micrograph images of materials tested and Figures S1-S5 described in the Methods Section. This material is available free of charge via the Internet at http://pubs.acs.org.

ACKNOWLEDGEMENTS

The authors are grateful for several sources of support. W. Miquelon, Chief Executive Officer of Jo-Ann Stores, Inc., generously donated many of the fabrics used in this study. Supplies and emergency laboratory access were enabled by Northeastern University through a COVID SEED grant. S. Lustig acknowledges partial funding from NSF/CBET grant 1604369 and support as a U.S. Army Research Lab joint faculty appointee. J. Biswakarma and D. Rana are supported by journeyman fellowships funded by the U.S. Army Research Laboratory through Oak Ridge Associated Universities/ Oak Ridge Institute for Science and Education. Other masks, respirators and samples were provided by M. Rosenblatt and Lahey Hospital and Medical Center. We would like to thank Ronald Willey, Interim Chair, Department of Chemical Engineering, and David Luzzi, Senior Vice Provost for Research Innovation and Development, Northeastern University for internal grants that enabled this work. Shoshanna Lustig provided the artwork in Figure 1.

REFERENCES

1 Eilperin, J.; Stein, J.; Butler, D.; Hamburger, T. U.S. Sent Millions of Face Masks to China Early This Year, Ignoring Pandemic Warning Signs, 18 April **2020**. [Online], https://www.washingtonpost.com/health/us-sent-millions-of-face-masks-to-china-early-this-year-ignoring-pandemic-warning-signs/2020/04/18/aaccf54a-7ff5-11ea-8013-1b6da0e4a2b7_story.html (accessed May 15, 2020).

2 Zhang, D.; Mansfield, E.; Pulver, D. V. U.S. Exported Millions in Masks and Ventilators Ahead of the Coronavirus Crisis," 2 April **2020**. [Online], https://www.usatoday.com/story/news/investigations/2020/04/02/us-exports-masks-ppe-china-surged-early-phase-coronavirus/5109747002/ (accessed May 5, 2020).

3 Rogers, K.; Spring, B. Personal Protective Equipment is in High Demand as Coronavirus Spreads, CNBC, 6 March **2020**. [Online], https://www.cnbc.com/2020/03/06/personal-protective-equipment-is-in-high-demand-as-coronavirus-spreads.html (accessed May 5, 2020).

4 Leung, N.H.; Chu, D. K.; Shiu, E. Y. ; Chan, K.-H.; McDevitt, J. J.; Hau, B. J.; Yen, H.-L.; Li, Y.; Ip, D. K.; Peiris, J. M.; Seto, W.-H.; Leung, G. M.; Milton, D. K.; Cowling, B. J. Respiratory Virus Shedding in Exhaled Breath and Efficacy of Face Masks [Online], *Nat. Med.* **2020** https://doi.org/10.1038/s41591-020-0843-2 (accessed May 5, 2020).

5 MacIntyre, C. R.; Cauchemez, S.; Dwyer, D. E.; Seale, H.; Cheung, P.; Browne, G.; Fasher, M.; Wood, J.; Gao, Z.; Booy, R.; Ferguson, N. Face Mask Use and Control of Respiratory Virus Transmission in Households, *Emerging Infect. Dis.* **2009**, *15*(2), 233-364.

6 Booth, C. M.; Clayton, M.; Crook, B.; Gawn, J. Effectiveness of Surgical Masks Against Influenza Bioaerosols," *J. Hosp. Infect.,* **2013**, *84*, 22-26.

7 Anfinrud, P.; Stadnytskyi, V.; Bax, C. E.; Bax, A. Visualizing Speech-Generated Oral Fluid Droplets with Laser Light Scattering, *N. England J. Med.,* **2020** [Online], https://www.nejm.org/doi/full/10.1056/NEJMc2007800, (accessed April 15, 2020).

8 Centers for Disease Control and Prevention, How to Protect Yourself & Others, April 24, **2020**. [Online], https://www.cdc.gov/coronavirus/2019-ncov/prevent-getting-sick/prevention.html (accessed May 15, 2020).

9 Abaluck, J. ; Chevalier, J. A.; Christakis, N. A.; Forman, H. P.; Kaplan, E. H.; Ko, A.; Vermund, S. H. The Case for Universal Cloth Mask Adoption and Policies to Increase Supply of Medical Masks for Health Workers," April 6, **2020**. [Online], https://papers.ssrn.com/sol3/papers.cfm?abstract_id=3567438 (accessed May 2020).

10 Centers for Disease Control, NIOSH Personal Protective Equipment Information (PPE-Info), November 20, **2015**. [Online]. https://wwwn.cdc.gov/PPEInfo/Standards/Info/ASTMF210011(2018) (accessed May 5, 2020).

11 Centers for Disease Control and Prevention, The National Institute for Occupational Safety and Health (NIOSH), May 25, **2018**. [Online], https://www.cdc.gov/niosh/index.htm (accessed May 6, 2020).

12 ASTM International, Standard Specification for Performance of Materials Used in Medical Face Masks, August 1, **2019**. [Online], https://www.astm.org/READINGLIBRARY/VIEW/F2100.html (accessed May 5, 2020).

13 ASTM International, Standard Test Method for Determining the Initial Efficiency of Materials Used in Medical Face Masks to Penetration by Particles Using Latex Spheres, June 1, **2017**. [Online], https://www.astm.org/READINGLIBRARY/VIEW/F2299.html (accessed May 5, 2020).

14 ASTM International, Standard Test Method for Evaluating the Bacterial Filtration Efficiency (BFE) of Medical Face Mask Materials, Using a Biological Aerosol of Staphlococcus aureus, July 1, **2019**. [Online], https://www.astm.org/READINGLIBRARY/VIEW/F2101.html (accessed May 5, 2020).

15 Han, Z.; Weng, W.; Huang, Q. Characterizations of Particle Size Distribution of the Droplets Exhaled by Sneeze. *J. R. Soc. Interface,* **2013**, *10*, 20130560.

16 Yang, S.; Lee, G. W.; Chen, C.-M.; Wu, C.-C.; Yu, K.-P. The Size and Concentration of Droplets Generated by Coughing in Human Subjects, *J. Aerosol Med.,* **2007**, *20*, 484-494.

17 Cascella, M.; Rajnick, M.; Cuomo, A.; Dulebohn, S. C.; Di Napoli, R. Features, Evaluation and Treatment of Coronavirus (COVID-19), April 6, **2020**. [Online], https://www.ncbi.nlm.nih.gov/books/NBK554776 (accessed May 5, 2020).

18 Rengasamy, S.; Eimer, B.; Shaffer, R. E. Simple Respiratory Protection- Evaluation of the Filtration Performance of Cloth Masks and Common Fabric Materials Against 20-1000 nm Size Particles, *Ann. Occup. Hyg.,* **2010**, *54*(7), 789-798.

19 Davies, A.; Thompson, K.-A.; Giri, K.; Kafatos, G.; Walker, J.; Bennett, A. Testing the Efficiency of Homemade Masks: Would They Protect in an Influenza Pandemic?, *Disaster Med. Public Health Prep.,* **2013**, *7*(4), 413-418.

20 Konda, A.; Prakash, A.; Moss, G. A.; Schmoldt, M.; Grant, G. D.; Guha, S. Aerosol Filtration Efficiency of Common Fabrics Used in Respiratory Cloth Masks, *ACS Nano,* **2020**, Article ASAP DOI: 10.1021/acsnano.0c03252 (accessed May 15, 2020).

21 Carroll, R. G. Pulmonary System," in *Elsevier's Integrated Physiology*, [Online], https://doi.org/10.1016/B978-0-323-04318-2.50016-9, Elsevier, 2007, pp. 99-115 (accessed May 15, 2020).

22 Dahneke, B. The Capture of Aerosol Particles by Surfaces, *J. Colloid Interface Sci.,* **1971**, *37*(2), 342-353.

23 ASTM International, ASTM F2100-19. Standard Specification for Performance of Materials Used in Medical Face Masks, 2019.

24 ASTM International, ASTM F2101-19, Standard Test Method for Evaluating the Bacterial Filtration Efficiency (BFE) of Medical Face Mask Materials, Using a Biological Aerosol of Staphylococcus aureus, 2019.

25 ASTM International, ASTM F2299M-03, Standard Test Method for Determining the Initial Efficiency of Materials Used in Medical Face Masks to Penetration by Particulates Using Latex Spheres, doi: 10.1520/F2299_F2299M-03R17, 2017.

26 Schneider, C.; Rasband, W.; Eliceiri, K. NIH Image to ImageJ: 25 Years of Image Analysis, *Nat. Methods,* **2012**, *9*, 671-675.

27 Marr, L.; Tang, J.; Van Mullekom, J.; Lakdawala, S. Mechanistic Insights into the Effect of Humidity on Airborne Influenza Virus Survival, Transmission and Incidence, *J.R. Soc. Interface,* **2019**, *16*(150), 20180298.